\documentclass[aps,showpacs,nofootinbib,pra,preprint,superscriptaddress,floatfix,11pt]{revtex4-1}
     
\usepackage{multirow}
\usepackage{amsfonts}
\usepackage{amsmath}
\usepackage{amssymb}
\usepackage{graphicx}
\usepackage{dcolumn}
\usepackage{times}
\usepackage{color}
\usepackage[a4paper]{geometry}
\usepackage{hhline}

\begin{document}

\title{Non-local divergence-free currents for the account of symmetries in two-dimensional wave scattering}

\author{M.~A.~Metaxas}
\affiliation{Department of Physics, University of Athens, GR-15784 Athens, Greece}

\author{P.~Schmelcher}
\email[]{pschmelc@physnet.uni-hamburg.de}
\affiliation{Zentrum f\"{u}r Optische Quantentechnologien, Universit\"{a}t Hamburg, Luruper Chaussee 149, 22761 Hamburg, Germany}
\affiliation{The Hamburg Centre for Ultrafast Imaging, Universit\"{a}t Hamburg, Luruper Chaussee 149, 22761 Hamburg, Germany}

\author{F.~K.~Diakonos}
\email[]{fdiakono@phys.uoa.gr}
\affiliation{Department of Physics, University of Athens, GR-15784 Athens, Greece}

\date{\today}

\begin{abstract}
We explore wave-mechanical scattering in two spatial dimensions assuming that the corresponding potential is invariant under linear symmetry transforms such as rotations, reflections and coordinate exchange. Usually the asymptotic scattering conditions do not respect the symmetries of the potential and there is no systematic way to predetermine their imprint on the scattered wave field. Here we show that symmetry induced, non-local, divergence-free currents can be a useful tool for the description of the consequences of symmetries on higher dimensional wave scattering, focusing on the two-dimensional case. The condition of a vanishing divergence of these non-local currents, being in one-to-one correspondence with the presence of a symmetry in the scattering potential, provides a systematic pathway to to take account if the symmetries in the scattering solution. It leads to a description of the scattering process which is valid in the entire space including the near field regime. Furthermore, we argue that the usual asymptotic representation of the scattering wave function does not account for insufficient account for a proper description of the underlying potential symmetries. Within our approach we derive symmetry induced conditions for the coefficients in the wave field expansion with respect to the angular momentum basis in two dimensions, which determine the transition probabilities between different angular momentum states.

\end{abstract}
\maketitle

\section{Introduction}\label{sec:1}
Symmetries play a significant role for the classification of eigenstates in wave mechanical systems ranging from acoustics and optics to quantum mechanics. In particular the impact of symmetries on wave scattering off a static potential in higher spatial dimensions has been explored in numerous works \cite{book1,book2,scatt_sym}. In most cases, when the potential decays rapidly at large radial distances $r$, the asymptotic form of the wave profile --within the far field approximation-- is used \cite{Asymp,Ramm}. However, as discussed in the recent literature \cite{Liu2014}, this asymptotic representation does not capture all details of the scattering process since it is not a solution of the underlying wave equation, unless the limit $r \to \infty$ is taken. This approach raises the question about the preservation of symmetry derived conditions, which have been posed before the limit $r \to \infty$ has been taken. An argument used in the application of symmetries to wave mechanical scattering systems is the independence of the form of the scattering wave function on the choice of the coordinate system \cite{Ramm}. However, this argument is weakened by the fact that the incoming wave, possessing a fixed asymptotic form, which is employed for the determination of the scattering wave profile, restricts this coordinate system independence. Thus, the description of the impact of symmetries in scattering systems is a subtle issue. 

In one-dimensional wave mechanical scattering it has been shown that the signature of the potential symmetry is carried by non-local currents constructed from the wave field at the original and the symmetry transformed spatial point. In the presence of a symmetry of the time-independent scattering potential, these currents turn out to be constant in space \cite{Kalozoumis2014,Kalozoumis2015}. This property holds even for the case when the potential symmetry is not globally valid, but appears only within finite spatial domains. In the latter case the constancy of the non-local currents holds only within the corresponding symmetry domain \cite{Kalozoumis2014,Kalozoumis2015}. The relation of these spatially constant currents to perfect transmission through potential landscapes possessing global or local symmetries, as well as their use as an order parameter to determine the corresponding phase diagrams in the context of PT symmetry, separating symmetric from non-symmetric scattering states, have been extensively studied in previous works \cite{LocSym}. Furthermore, recently it has been shown that these currents are obtained, as a special case, from a more general continuity equation involving currents constructed from wave fields satisfying two different wave equations \cite{Diakonos2019}. Even more, as shown in \cite{Diakonos2019,Schmelcher2017, Spourdalakis2016} such currents can be defined for arbitrary spatial dimensions $d$ and number of degrees of freedom $N$. The only significant change for $d > 1$ is that these generalized currents are not spatially constant any more but they possess a vanishing divergence. Thus, the aforementioned generalized currents in dimensions $d > 1$ when applied to scattering in a locally or globally symmetric potential become, in the sense mentioned previously, non-local divergence-free currents.

Within the framework of wave mechanical scattering in higher dimensions it is natural to ask if these, symmetry related, non-local currents could be useful for the description of the impact of the potentials' symmetries on the scattered wave field. We consider this question in the present work, focusing on two-dimensional quantum mechanical scattering. As stated above, the link between symmetries of the scattering potential and the resulting structure of the scattering wave function is blurred by the fact that the boundary conditions implied by the scattering set-up break, in general, some or all of the symmetries of the potential. As a consequence, the exact wave function ceases to be an eigenstate of the associated symmetry transforms and therefore its structure is not directly connected to the symmetries of the potential. Nevertheless, the condition of zero divergence for the suitably constructed, symmetry induced, non-local currents discussed above, is still valid. Since these currents contain the exact scattering wave field, their vanishing divergence can be considered as a remnant of the potentials' symmetry, which is broken at the level of the wave field, due to the boundary conditions. In this sense the symmetry induced divergence-free, non-local currents provide the missing link which is necessary for the identification of the impact of the potentials' symmetries on the scattering field. To make practical use of this property, we will first expand the scattering solution in the polar basis of the free Schr\"{o}dinger Hamiltonian in two dimensions and insert this expansion into the expression defining the non-local currents. Due to the symmetry breaking, an infinite superposition of radial modes corresponds to each angular momentum state. Thus, the matrix determining the radial part in this expansion ceases to be diagonal. In this description, the symmetries of the potential, expressed by the vanishing divergence of these currents, lead to conditions on the matrix elements of the radial part, providing us with a determination of the symmetry induced structure of the scattered state, based on first principles. In fact, these conditions lead to constraints among the transition probabilities between the different radial modes.

Our work is organized as follows: in section II we present a short derivation of the symmetry induced non-local currents for $d > 1$, as well as their polar form assuming $d=2$. Additionally, we show that these currents can be generated by a scalar potential, which exhibits the associated symmetry properties. In section III we discuss the inconsistency of the usual asymptotic ansatz for the wave field in $2d$ quantum scattering, with respect to the vanishing of the divergence of the symmetry induced non-local currents derived in section II. In section IV we employ the non-local currents to derive conditions for the stationary scattering dynamics, imposed by the presence of a global symmetry in the time-independent scattering potential. Finally, in section V we give our concluding remarks and shortly discuss possible extensions of our work.

\section{Derivation of divergence-free currents}\label{sec:2}

Since a multitude of physical experiments measure the outcome of scattering events, scattering theory in arbitrary spatial dimensions is an important and still very active research field. Here we will study quantum scattering in two dimensions adopting the viewpoint of divergence-free non-local currents implied by the symmetries of the scattering potential. Our first task is to derive these currents in Cartesian coordinates assuming the invariance of the potential under specific geometric transformations implying certain symmetries. 

In the following we will focus on geometric transformations $\mathbf{t}: \mathbf{r} \mapsto \bar{\mathbf{r}}$ in $\mathbb{R}^2$ with $\bar{\mathbf{r}}=\mathbf{t}(\mathbf{r})$, which leave the magnitude of the vector $\mathbf{r}$ invariant and in addition they do not change the location of the origin $O=(0,0)$ of the reference system. However, the procedure we will develop, holds for more general cases, such as spatial translations or dilatations, as well. Specifically, we will consider the following transformations:

\begin{itemize}

\item Reflection with respect to the $y$ axis, $\left( {x,y} \right) \mapsto \left( {\bar x,\bar y} \right) = \left( { - x,y} \right)$ or in polar form $\left(\bar{r},\bar{\phi}\right)=\left(r,\pi - \phi\right)$, with $r$ being the radial distance and $\phi$ the azimuthal angle.

\item Reflection with respect to $x$ axis, $\left( {x,y} \right) \mapsto \left( {\bar x,\bar y} \right) = \left( {  x,-y} \right)$ or in polar form $\left(\bar{r},\bar{\phi}\right)=\left(r,-\phi\right)$. 

\item Rotation in the $x$-$y$ plane  through a finite angle $\phi_0$, $\left( {x,y} \right) \mapsto \left( {\bar x,\bar y} \right)$ with \\
$\left( {\bar x,\bar y} \right)= \left( x \cos \phi_0  - y \sin \phi_0,  x\sin \phi_0  + y\cos \phi_0  \right)$ or in polar form $\left(\bar{r},\bar{\phi}\right)=\left(r,\phi + \phi_0 \right)$. As a special case of this transform we obtain, for $\phi_0=\pi$, the inversion with respect to the origin.

\item Exchange between $x$ and $y$, $\left( {x,y} \right) \mapsto \left( {\bar x,\bar y} \right) = \left( { y,x} \right)$ or in polar form $\left(\bar{r},\bar{\phi}\right)=\left(r,{\pi \over 2}-\phi\right)$.    
\end{itemize} 

All these transforms can be summarized in a single relation, as follows
\begin{equation}
\label{eq:1}
\bar{r}=r, \bar{\phi}=s \pm \phi~~~~~~;~~~~~~s=0,~{\pi \over 2},~\pi,~\phi_0.
\end{equation}

\noindent
The stationary Schr\"odinger equation for a fixed energy eigenvalue $E$, in coordinate representation, is written as  
\begin{equation}
\label{eq:2}
\hat{H} \Psi(\mathbf{r}) = E \Psi(\mathbf{r})~~\mathrm{with}~~\hat{H}= -\frac{\hbar^2}{2 m_p} \mathbf{\nabla}^2 + V(\mathbf{r})
\end{equation} 
where $\hat{H}$ is the Hamilton operator and $m_p$ is the mass of the quantum particle. Denoting as $\hat{T}$ the Hilbert space operator corresponding to the $\mathbb{R}^2$-transform $\mathbf{t}$, we find that -- since $\mathbf{t}$ does not change the magnitude of a vector in $\mathbb{R}^2$ -- the Laplacian operator $\mathbf{\nabla}^2$ is invariant under the action of $\hat{T}$. We further assume that the potential $V(\mathbf{r})$ is also invariant under $\hat{T}$ which in turn implies that $\hat{T} \hat{H} \hat{T}^{-1} = \hat{H}$, i.e. the Hamilton operator $\hat{H}$ is invariant under the action of $\hat{T}$. Under this conditions, applying the transform $\hat{T}$ to Eq.~(\ref{eq:2}) we obtain
\begin{equation}
\label{eq:3}
\hat{H} \Psi(\mathbf{t}(\mathbf{r})) = E \Psi(\mathbf{t}(\mathbf{r}))
\end{equation} 
with $\Psi(\mathbf{t}(\mathbf{r})) = \hat{T} \Psi(\mathbf{r})$. We multiply Eq.~(\ref{eq:2}) with $\Psi(\bar{\mathbf{r}})$ and Eq.~(\ref{eq:3}) with $\Psi(\mathbf{r})$ using the notation $\mathbf{\bar{r}}=\mathbf{t}(\mathbf{r})$. Subtracting the resulting expressions and taking into account that the potential is invariant under the transform $\mathbf{r} \mapsto \mathbf{\bar{r}}$, i.e. $ V(\mathbf{r})=V(\mathbf{\bar{r}})$ we find
\begin{equation}
\label{eq:4}
\Psi(\mathbf{\bar{r}}){\nabla^2}\Psi(\mathbf{r}) - \Psi(\mathbf{r}){\nabla^2}\Psi(\mathbf{\bar{r}})=0
\end{equation}
Eq.~(\ref{eq:4}) can be written as
\begin{equation}
\label{eq:5}
\mathbf{\nabla} \cdot \left( \Psi(\mathbf{\bar{r}}){\mathbf{\nabla}}\Psi(\mathbf{r}) - \Psi(\mathbf{r}){\mathbf{\nabla}}\Psi(\mathbf{\bar{r}}) \right)=0 
\end{equation}
leading to the definition of the non-local, divergent-free stationary current
$\mathbf{Q}(\mathbf{r},\mathbf{\bar{r}})$ (NLC), induced by the symmetry of the scattering potential $V(\mathbf{r})$
\begin{equation}
\label{eq:6}
\mathbf{Q}(\mathbf{r},\mathbf{\bar{r}})=\frac{1}{2 i} \left( \Psi(\mathbf{\bar{r}}){\mathbf{\nabla}}\Psi(\mathbf{r}) - \Psi(\mathbf{r}){\mathbf{\nabla}}\Psi(\mathbf{\bar{r}}) \right)
\end{equation}
Assuming that the scattering potential is real, one can use the complex conjugate of Eq.~(\ref{eq:2}) and repeat the previously described procedure, to define an additional non-local, divergence-free, symmetry induced current $\mathbf{\tilde{Q}}$ given as
\begin{equation}
\label{eq:7}
\mathbf{\tilde{Q}}(\mathbf{r},\mathbf{\bar{r}})=\frac{1}{2 i} \left( \Psi(\mathbf{\bar{r}}){\mathbf{\nabla}}\Psi^*(\mathbf{r}) - \Psi^*(\mathbf{r}){\mathbf{\nabla}}\Psi(\mathbf{\bar{r}}) \right)
\end{equation}
It is straightforward to show that the two non-local currents are not independent, but they are related to each other through the constraint
\begin{equation}
\label{eq:8}
\mathbf{Q}(\mathbf{r},\mathbf{\bar{r}}) \cdot \mathbf{Q}^*(\mathbf{r},\mathbf{\bar{r}})-\mathbf{\tilde{Q}}(\mathbf{r},\mathbf{\bar{r}}) \cdot \mathbf{\tilde{Q}}^*(\mathbf{r},\mathbf{\bar{r}})=\mathbf{J}(\mathbf{r}) \cdot \mathbf{\bar{J}}^*(\mathbf{\bar{r}})
\end{equation}
where 
\begin{equation}
\label{eq:9}
\mathbf{J}(\mathbf{r})=\frac{1}{2i} \left ( \Psi^*(\mathbf{r}){\mathbf{\nabla}}\Psi(\mathbf{r}) - \Psi(\mathbf{r}){\mathbf{\nabla}}\Psi^*(\mathbf{r}) \right) 
\end{equation}
is the usual, local probability current for a stationary energy eigenstate and 
\begin{equation}
\label{eq:10}
\mathbf{\bar{J}}(\bar{\mathbf{r}})=\frac{1}{2i} \left ( \Psi^*(\mathbf{\bar{r}}){\mathbf{\nabla}}\Psi(\mathbf{\bar{r}}) - \Psi(\mathbf{\bar{r}}){\mathbf{\nabla}}\Psi^*(\mathbf{\bar{r}}) \right)
\end{equation}
is the probability current evaluated using the wave field $\Psi(\mathbf{\bar{r}})$. Notice that in Eq.~(\ref{eq:10}) the nabla operator is expressed in the variable $\mathbf{r}$ and not in $\mathbf{\bar{r}}$, therefore $\mathbf{\bar{J}}(\bar{\mathbf{r}}) \neq \mathbf{J}(\mathbf{\bar{r}})$. The condition given in Eq.~(\ref{eq:8}) is the higher dimensional extension of a similar relation holding in the one-dimensional case \cite{Kalozoumis2015}. 

One can show that the NLC $\mathbf{Q}(\mathbf{r},\mathbf{\bar{r}})$ vanishes whenever the state $\Psi(\mathbf{r})$ is an eigenstate of the operator $\hat{T}$ realizing the transformation $\mathbf{\bar{r}}=\mathbf{t}(\mathbf{r})$ in Hilbert space. $\mathbf{Q}$ can therefore be used as an indicator (order parameter) for the symmetry breaking related to this transform, similarly to the one-dimensional case \cite{Kalozoumis2015}. Usually, in a scattering process, symmetry is broken by the incident plane wave component $\Psi_{in}(\mathbf{r})$ and therefore, in general, the NLC $\mathbf{Q}$ is different from zero. In the present work, our aim is to show how the NLC $\mathbf{Q}$ can be used to gain information for a scattering state in two spatial dimensions. The path we will follow can, in principle, be transferred also to the three-dimensional case. However, a detailed analysis of the latter involves a modified analytical treatment and goes beyond the scope of the present study. It will be investigated in a forthcoming work \cite{Metaxas}. 

It is useful to express the components of $\mathbf{Q}$ in Cartesian and polar coordinates. Obviously, the Cartesian components of $\mathbf{Q}$ are
\begin{equation}
\label{eq:11}
\mathbf{Q} = \frac{1}{2 i} \left( \mathit{\Psi}(\bar{x},\bar{y}){\partial_x}\mathit{\Psi}(x,y) - \mathit{\Psi}(x,y){\partial_x}\mathit{\Psi}(\bar{x},\bar{y}),\mathit{\Psi}(\bar{x},\bar{y}){\partial_y}\mathit{\Psi}(x,y) - \mathit{\Psi}(x,y){\partial_y}\mathit{\Psi}(\bar{x},\bar{y}) \right) 
\end{equation}
while the corresponding polar components are
\begin{eqnarray}
\label{eq:12}
Q_r &=& {1 \over 2i} \left(\Psi(\bar{r},\bar{\phi}){\partial_r}\Psi(r,\phi) - \Psi(r,\phi){\partial_r}\Psi(\bar{r},\bar{\phi})\right) \nonumber \\
Q_{\phi} &=& {1 \over 2i r} \left(\Psi(\bar{r},\bar{\phi}){\partial_{\phi}}\Psi(r,\phi) - \Psi(r,\phi){\partial_{\phi}}\Psi(\bar{r},\bar{\phi})\right)
\end{eqnarray}
where we have used the symbols $\mathit{\Psi}$, $\Psi$ for the wave function in Cartesian and polar coordinates respectively. Since $\mathbf{Q}$ is a complex two-dimensional vector, Cartesian and polar components are related in the same way as a $\mathbb{R}^2$-vector
\begin{eqnarray}
\label{eq:13}
Q_x &=& Q_r \cos \phi  - Q_{\phi} \sin \phi \nonumber \\
Q_y &=& Q_r \sin \phi  + Q_{\phi}\cos \phi
\end{eqnarray}

It is interesting to notice that $\mathbf{Q}$ in two dimensions, possessing a vanishing divergence, can be represented in terms of the gradient of a complex scalar "potential" $\Phi(\mathbf{r})$  \cite{Barbarosie2011}:
\begin{equation}
\mathbf{Q}=\hat{R}^{-1} \mathbf{\nabla} \Phi(\mathbf{r})~~~~;~~~~\hat{R}=\left[ \begin{array}{cc} 0 & -1 \\ 1 & 0 \\ \end{array} \right]
\label{eq:14}
\end{equation}
which, in general, carries the information of the symmetry transform. A relation similar to Eq.~(\ref{eq:14}) holds also for $\mathbf{\tilde{Q}}$ with a corresponding potential $\tilde{\Phi}(\mathbf{r})$. In fact, it is possible to invert Eq.~(\ref{eq:14}) and determine the generating potential $\Phi(\mathbf{r})$ in terms of the components of the current $\mathbf{Q}$. To this end we use the polar form of $\mathbf{Q}$
\begin{equation}
\mathbf{Q} = Q_r(r,\phi,\bar{r},\bar{\phi}) \mathbf{e_r} + Q_{\phi}(r,\phi,\bar{r},\bar{\phi})\mathbf{e_{\phi}}
\label{eq:15}
\end{equation}
where $\mathbf{e_r}$, $\mathbf{e_{\phi}}$ are the unit vectors in the corresponding directions. Since in the polar representation $(\bar{r},\bar{\phi})$ can be expressed in terms of $(r,\phi)$ one can expand $\mathbf{Q}$ solely in terms of the degrees of freedom $(r,\phi)$ as follows
\begin{equation}
\mathbf{Q}=\displaystyle{\sum_{m=-\infty}^{\infty}} \left( Q_{m,r}(r)
\mathbf{e_r} + Q_{m,\phi}(r) \mathbf{e_{\phi}} \right) e^{i m \phi}
\label{eq:16}
\end{equation}
Employing the expansion (\ref{eq:16}) and the vanishing of the divergence of $\mathbf{Q}$ we obtain
\begin{equation}
\frac{d(r Q_{m,r})(r)}{dr}=-i m Q_{m,\phi}(r)~~~~~~~;~~~~~~~\forall~ m \in \mathbb{Z}
\label{eq:17}
\end{equation}
Setting $m=0$ in  Eq.~(\ref{eq:17}) we find the general condition
\begin{equation}
Q_{0,r}(r)=\frac{C}{r}
\label{eq:18}
\end{equation}
with $C$ a complex constant. From the relations
\begin{equation}
Q_r=\frac{1}{r}\frac{\partial \Phi}{\partial \phi}~~~;~~~Q_{\phi}=-\frac{\partial \Phi}{\partial r}
\label{eq:19}
\end{equation}
implied by Eq.~(\ref{eq:14}) we obtain
\begin{equation}
\Phi(r,\phi)=C \phi - i r \displaystyle{\sum_{\stackrel{m=-\infty}{m \neq 0}}^{\infty}} \frac{1}{m} Q_{m,r}(r) e^{i m \phi}
\label{eq:20}
\end{equation}
For $\Phi(r,\phi)$ to be single-valued ($\Phi(r,\phi)=\Phi(r,\phi+ 2 \pi)$), the constant $C$ has to be zero, meaning that in general $Q_{0,r}(r)=0$. One can proceed further using the general representation of the scattering wave function in polar coordinates
\begin{equation}
\label{eq:21}
\Psi(r,\phi)=\displaystyle{\sum_{m=-\infty}^{\infty}} g_m(r) e^{i m \phi}
\end{equation}
For the class of transformations given in Eq.~(\ref{eq:1}), $Q_{m,r}(r)$ can be written as
\begin{equation}
\label{eq:22}
Q_{m,r}(r)=\frac{1}{2i} \displaystyle{\sum_{n=-\infty}^{\infty}} \left( g_n(r) \frac{d g_{m \mp n}(r)}{d r} - g_{m \mp n}(r) \frac{d g_n(r)}{d r} \right) e^{i n s}
\end{equation} 
which in turn leads to the following expression for the generating scalar potential $\Phi(r,\phi)$
\begin{eqnarray}
\label{eq:23}
\Phi(r,\phi)=&-&\frac{r}{2} \displaystyle{\sum_{m=1}^{\infty} \frac{1}{m}  \sum_{n=-\infty}^{\infty}} e^{i n s} \left[ g_n(r) \left(\frac{d g_{m \mp n}(r)}{d r} e^{i m \phi} - \frac{d g_{-m \mp n}(r)}{d r} e^{-i m \phi} \right) 
\right. \nonumber \\
&-&\left. \frac{d g_n(r)}{d r} \left(g_{m \mp n}(r) e^{i m \phi} - g_{-m \mp n}(r) e^{- i m \phi} \right) \right]
\end{eqnarray}

It is useful to present an illustration of the potential $\Phi(r,\phi)$, for various symmetry transforms, in a specific example. Scattering off a hard circular disc potential is perfectly suited for this, since: (i) it is easily solved analytically and (ii) it possesses all the symmetries listed above. In particular, for this potential the rotational symmetry holds for any angle $\phi_0$. Thus, in the following we will employ this example for illustrative reasons, despite the fact that the developed theoretical tools are valid for the most general case of two-dimensional scattering. The wave function of the hard disk scattering problem is given as
\begin{equation}
\label{eq:24}
\Psi(r,\phi)=\displaystyle{\sum_{m=-\infty}^{\infty}} i^m \left( J_m(k r) - \frac{J_m(k R)}{H^{(+)}_m(kR)} H^{(+)}_m(k r) \right) e^{i m \phi}~~~;~~~r \geq R
\end{equation}
where $J_m$ and $H^{(\pm)}_m$ are the $m$-th order Bessel and complex Hankel functions respectively \cite{Abramowitz1973}, while $R$ is the disk radius. In Eq.~(\ref{eq:24}) $k$ is the wave vector of the incoming wave, assumed to lie in the positive $x$ direction. Thus, the function $g_m(r)$ for the hard disk potential has the form
\begin{equation}
\label{eq:25}
g_m(r)=e^{i m \pi /2} \left( J_m(k r)-\frac{J_m(k R)}{H^{(+)}_m(k R)} H^{(+)}_m(k r) \right)
\end{equation}
Inserting Eq.~(\ref{eq:25}) into Eq.~(\ref{eq:23}) we can calculate the generating potential $\Phi(r,\phi)$ for the hard disk scattering.
The case of reflections with respect to the $x$-axis is trivial, since, if the incoming wave vector is along the $x$-direction, it is an exact symmetry of the wave field. Therefore, the divergence-free current $\mathbf{Q}$ and its generating potential $\Phi(r,\phi)$ vanish. Next we consider the scalar potential generating the NLC $\mathbf{Q}$ related to the invariance of the scattering potential under reflections with respect to the $y$-axis. In Fig.~1 we show the real and imaginary part of ${\Phi(r,\phi) \over r}$ in (a) and (b) respectively, while in Fig.~2 we show the corresponding generating potential of the NLC for the transform $\bar{\phi}=\phi+{\pi \over 2}$ (rotation by ${\pi \over 2}$). In each case, the underlying symmetry transform is encoded in the extrema of $\Phi(r,\phi)$ where the NLC locally vanish. In addition, the potential in Fig.~1 is an odd function of $\phi$ indicating that the terms proportional to $\cos m \phi$ in Eq.~(\ref{eq:23}), for the hard circular disk potential where $g_m(r)=g_{-m}(r)$, vanish.  In contrast, the potential in Fig.~2 is locally parity symmetric, appearing as an even function around the values $-{\pi \over 4}$ and ${3 \pi \over 4}$. Furthermore, it can be shown that the generating potential for the transform $\bar{\phi}=\phi + \pi$ (reflection with respect to the origin) possesses exactly the same form as $\Phi(r,\phi)$ for the reflection with respect to the $y$-axis, shown in Fig.~1.

\begin{figure}[tbp]
\centering
\includegraphics[width=1.0\textwidth]{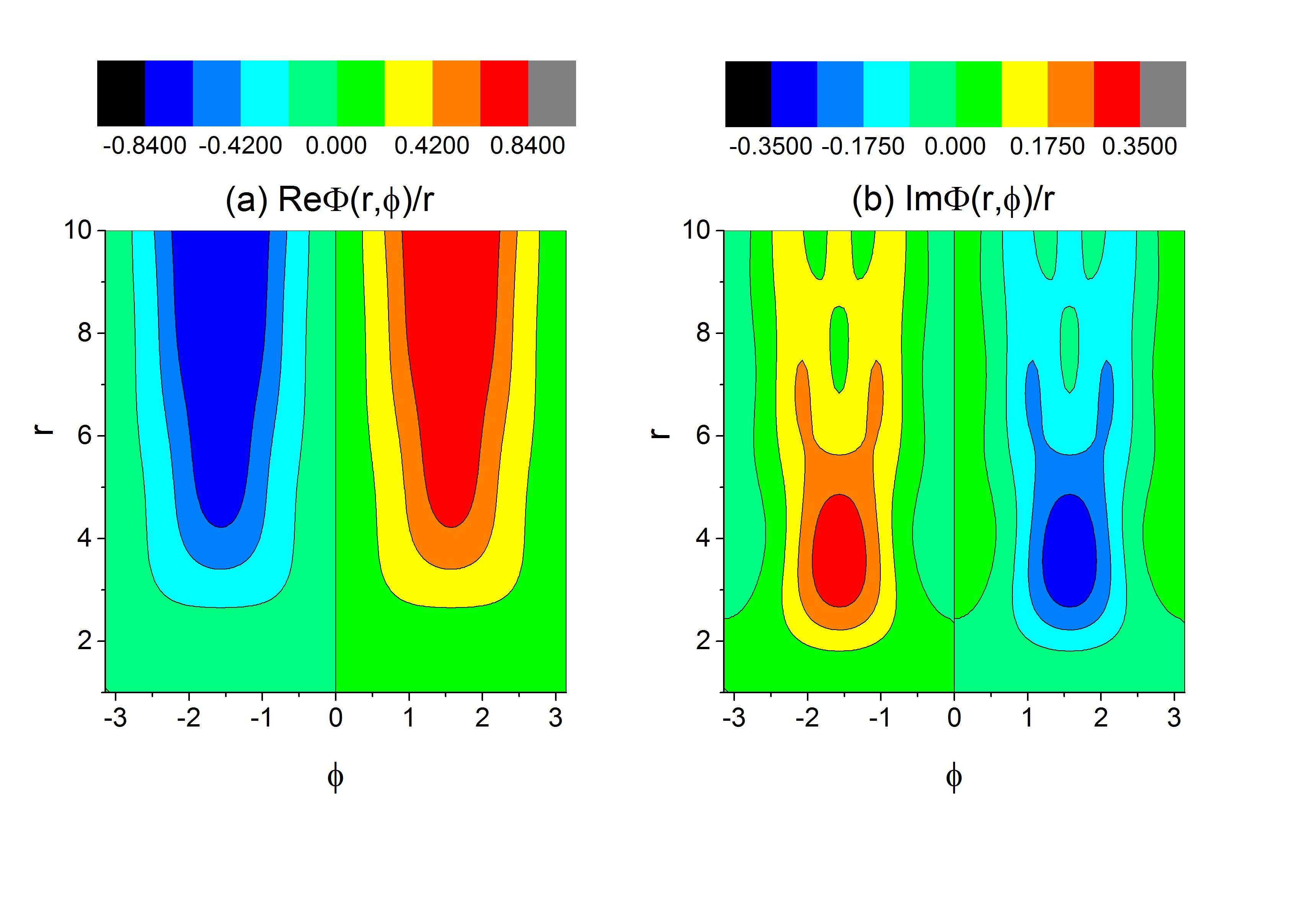}
\caption{The real (a) and complex (b) part of the scalar potential $\Phi(r,\phi)$ generating the non-local current $\mathbf{Q}$ for scattering off a hard disk potential, with $k R =1$ ($R$ being the disk radius and $k$ is the incoming wave vector magnitude), in the case of reflection with respect to the $y$-axis. The corresponding scalar potential for rotation by $\pi$ possesses exactly the same form.}
\label{fig:f1}
\end{figure} 

\begin{figure}[tbp]
\centering
\includegraphics[width=1.0\textwidth]{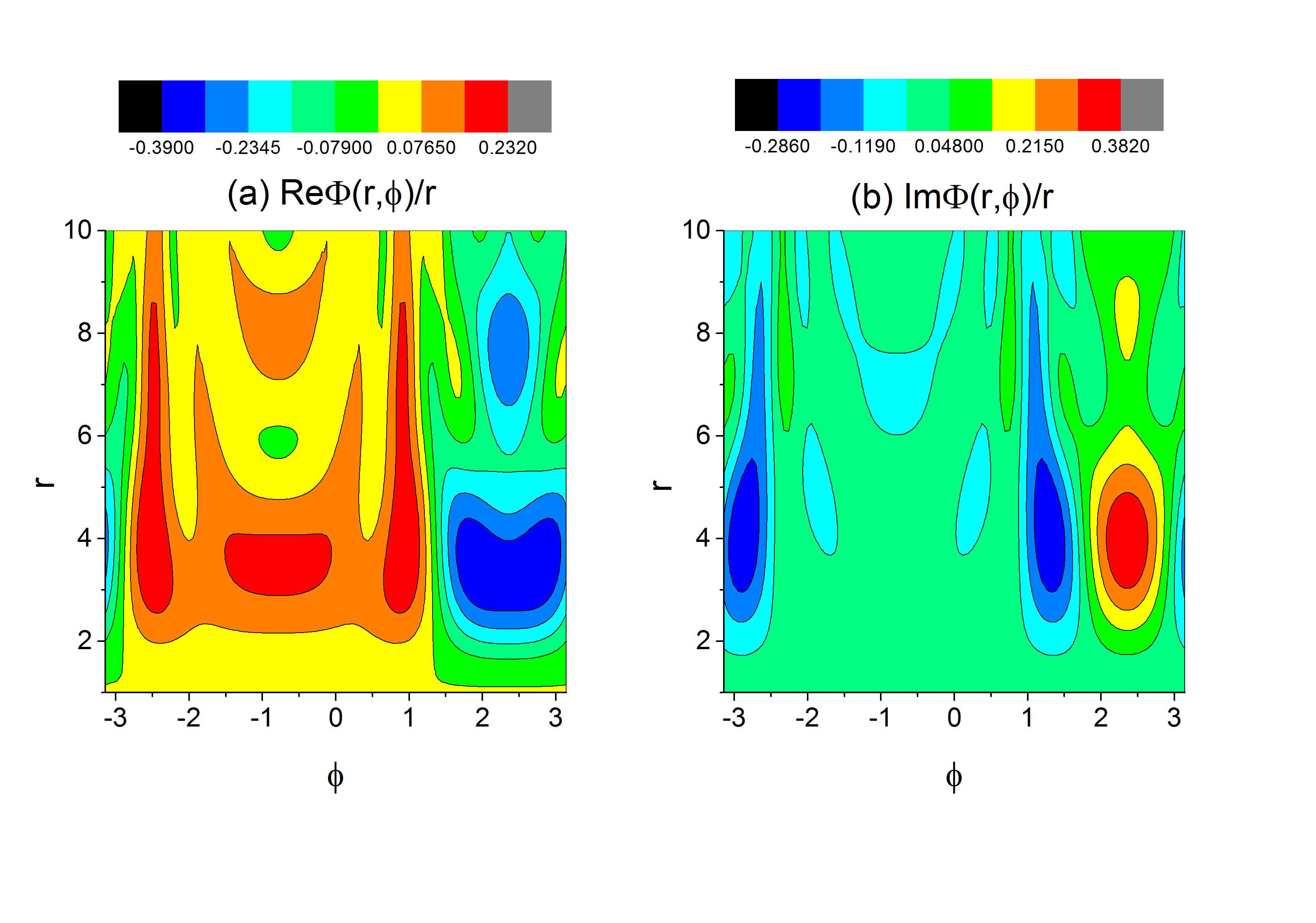}
\caption{The real (a) and complex (b) part of the scalar potential $\Phi(r,\phi)$ generating the NLC $\mathbf{Q}$ for scattering off a hard disk potential, with $k R =1$ ($R$ is the disk radius, $k$ is the incoming wave vector magnitude), in the case of rotation by ${\pi \over 2}$.}
\label{fig:f2}
\end{figure} 

\section{Asymptotics of the two-dimensional quantum scattering}
\label{sec:3}

In scattering problems the potential $V(\mathbf{r})$ decays to zero for the radial distance approaching infinity. Here we will assume that the decay, for sufficiently large $r$, is faster than ${1 \over r}$. All subsequent considerations and the theoretical framework which will be developed, refer to potentials of this form. In this case, according to the standard treatment of quantum 2-d scattering, for $r \to \infty$ the total wave function attains the form 
\begin{equation}
\label{eq:26}
\Psi_{as}\left(r,\phi,\phi_k \right)\mathop  \to \limits_{r \to \infty } e^{i\mathbf{k} \cdot \mathbf{r}} + \bar{F}\left(\phi,\phi_k,k\right)\frac{e^{ikr}}{\sqrt{k r}}
\end{equation}
with $k=\vert \mathbf{k} \vert$ and $\phi_k$ being the angle between the vectors $\mathbf{k}$ and $\mathbf{r}$. The first term on the r.h.s. of Eq.~(\ref{eq:26}) is the incoming wave $\Psi_{in}$ while the second term is the outgoing one $\Psi_{out}$. Without loss of generality we can make the assumption $\phi_k=0$ choosing the wave vector of the incoming wave along the $x$-axis. The angular part $F\left(\phi,k\right)=\bar{F}\left(\phi,0,k\right)$ can be expanded in partial waves with definite angular momentum $m \hbar$ ($m \in \mathcal{Z}$)
\begin{equation}
\label{eq:27}
F\left(\phi,k\right)=\displaystyle{\sum_{m=-\infty}^{+\infty}} f_m(k) e^{i m \phi}
\end{equation}
There are some important shortcomings of the asymptotic expansion (\ref{eq:26}) w.r.t. the treatment of symmetries. To reveal this, let us focus on finite support scattering potentials shown in Fig.~3. For this subclass of scattering potentials $V(\mathbf{r})$ is different from zero only within a finite subdomain $\mathcal{D}$ of $\mathcal{R}^2$ and therefore the free Schr\"{o}dinger equation (FSE) is obeyed for $\mathbf{r} \notin \mathcal{D}$. This property enables some analytical calculations. Notice that the hard disk potential used to calculate the generating potential in Figs.~1(a,b) and 2(a,b) is a special case of the class of finite support potentials.

\begin{figure}[tbp]
\centering
\includegraphics[width=0.75\textwidth]{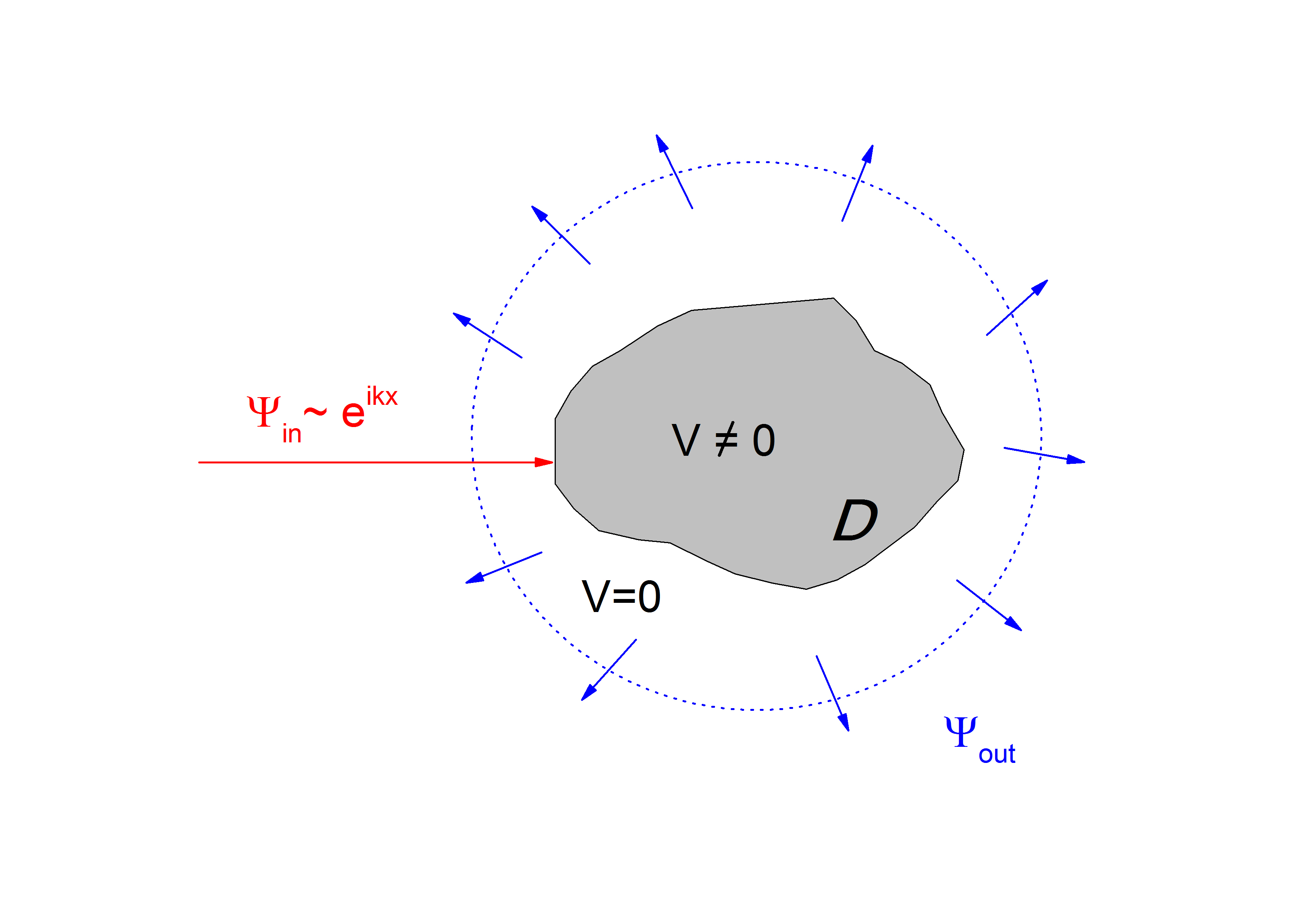}
\caption{A typical quantum scattering problem off a potential of finite support in two spatial dimensions. Within the shaded region the potential is different from zero while outside of this region it vanishes. The plane wave $\Psi_{in}$ (red arrow) with fixed wave vector $\mathbf{k}$ is incoming, while $\Psi_{out}$ is the outgoing solution of the corresponding Schr\"{o}dinger equation. The total wave field at energy $E=\frac{\hbar^2 \mathbf{k}^2}{2 m_p}$ is $\Psi=\Psi_{in}+\Psi_{out}$. }
\label{fig:f3}
\end{figure} 

Clearly, for scattering potentials of the type shown in Fig.~3, both $\Psi_{in}$ and $\Psi_{out}$ should satisfy the FSE. However, inserting the representation in Eq.~(\ref{eq:26}) for the asymptotic wave field $\Psi_{as}(\mathbf{r})$ into the 2-d FSE we find
\begin{equation}
\label{eq:28}
{\nabla^2}\Psi_{as}(\mathbf{r}) + \frac{2m_p E}{\hbar ^2}\Psi_{as}(\mathbf{r}) = 
e^{ikr} r^{-{5 \over 2}}\left( {\frac{{{\partial ^2}F(\phi,k)}}{{\partial {\phi ^2}}} + \frac{1}{4}F(\phi,k)} \right) + {\large{O}}(r^{-3})
\end{equation}
The term on the r.h.s. of Eq.~(\ref{eq:28}) is different from zero (for $r < \infty$) unless $F(\phi,k)=A(k)e^{i {\phi \over 2}}$, which is generally not the case. This dictates that the consistency of the description of the scattering wave is recovered only after taking the limit $r \to \infty$. But in this far field limit the finite support potential behaves as point-like and therefore specific information concerning its symmetries, contained in the domain $\mathcal{D}$, can be lost. 

Of course, this property is blurred if the potential decays to zero smoothly for $r \to \infty$, since in this case the ansatz in Eq.~(\ref{eq:26}) is a priori valid only at infinite distance from the origin while the solution at finite $\mathbf{r}$ is in general not known analytically. In short, for scattering in finite support potentials the asymptotic ansatz (\ref{eq:26}) does not provide a consistent description and cannot capture important properties, such as the invariance of the potential shape under specific coordinate transforms. This is also reflected in the divergence of the NLC $\mathbf{Q}_{as}$ (or $\mathbf{\tilde{Q}}_{as}$), calculated through $\Psi_{as}(\mathbf{r})$, which possesses a behaviour of the order of $r^{-\frac{5}{2}}$ for $r \to \infty$. Notice that the ansatz (\ref{eq:26}) contains a mixture of terms: the incoming component $\Psi_{in}(\mathbf{r})$ which satisfies the FSE, being an infinite sum in powers of ${1 \over \sqrt{r}}$, and the outgoing component $\Psi_{out}(\mathbf{r})$ which is not an exact solution of the FSE and contains only the leading term in ${1 \over \sqrt{r}}$ expansion. This property is the reason why both, the FSE as well as the zero-divergence condition of the NLCs are violated by $\Psi_{as}(\mathbf{r})$ with terms which are of the same order ($\large{O}(r^{-\frac{5}{2}})$) in the ${1 \over \sqrt{r}}$ expansion. Since both $\Psi_{in}(\mathbf{r})$ and $\Psi_{out}(\mathbf{r})$ should in principle fulfill the FSE for scattering in potentials with finite support, one could define symmetry induced non-local divergence-free currents $\mathbf{Q}_{in}$ and $\mathbf{Q}_{out}$ respectively. Clearly, the current $\mathbf{Q}_{in}$ possesses a vanishing divergence since $\Psi_{in}(\mathbf{r})$ satisfies the FSE while the divergence of the current $\mathbf{Q}_{out}$ turns out to be of the order of $\large{O}(r^{-3})$ in contrast to the violation of the FSE by $\Psi_{as}(\mathbf{r})$ which is of the order of $\large{O}(r^{-\frac{5}{2}})$. Thus, the mixing of different order terms in $\Psi_{as}(\mathrm{r})$ leads to stronger violation of the zero-divergence condition for the associated current $\mathbf{Q}_{as}$.

For scattering potentials of finite support with rotational symmetry one can replace the representation (\ref{eq:26}) by an exact one, employing the complex Hankel functions $H^{(\pm)}_m(x)$ \cite{Abramowitz1973} which solve the 2-d FSE. In fact, in this case, the scattering wave $\Psi_{out}(\mathbf{r})$ is given by
\begin{equation}
\label{eq:29}
\Psi_{out}(\mathbf{r})=\displaystyle{\sum_{m=-\infty}^{\infty}} \left( A_m H^{(+)}_m(k r) + B_m H^{(-)}_m(kr) \right) e^{i m \phi}
\end{equation}
with $A_m$, $B_m$ $\in~\mathbb{C}$. Eq.~(\ref{eq:29}) is compatible with the expression (\ref{eq:26}), involving outgoing asymptotic conditions, if we set $B_m=0$ for all $m$ and consider the leading term in the ${1 \over r}$ expansion of the remaining Hankel functions. Using Eq.~(\ref{eq:29}) the total wave field attains the form 
\begin{equation}
\label{eq:30}
\Psi(\mathbf{r})=e^{i\mathbf{k} \cdot \mathbf{r}} + \displaystyle{\sum_{m=-\infty}^{\infty}} A_m H^{(+)}_m(k r)  e^{i m \phi}
\end{equation}
outside of the region $\mathcal{D}$. It is straightforward to show that the symmetry induced non-local current $\mathbf{Q}$, containing the wave field $\Psi(\mathbf{r})$ in Eq.~(\ref{eq:30}) and its transformed counterpart $\Psi(\bar{\mathbf{r}})$, possess a vanishing divergence. Thus, through the ansatz (\ref{eq:29}) the inconsistent treatment of symmetries through the asymptotic form (\ref{eq:26}) is repaired. However, Eqs.~(\ref{eq:29},\ref{eq:30}) are only valid outside of the domain $\mathcal{D}$ for finite support potentials with rotational symmetry. In the next section we will develop a general scheme allowing a consistent incorporation of the symmetries in the description of scattering for the entire class of potentials decaying to zero faster than ${1 \over r}$. 

\section{Non-local divergence-free currents and symmetry induced constraints}\label{sec:4}

In this section we will explore the implications of the condition $\mathbf{\nabla}  \cdot \mathbf{Q} = 0$ on the form of the scattering field without any additional assumption related to the corresponding asymptotic behaviour. Our aim is to derive symmetry induced properties providing a wave field representation which is valid also in the near field regime. To this end, we employ the general expansion of the scattering wave function in 2-d 
\begin{equation}
\label{eq:31}
\Psi (\mathbf{r}) = \displaystyle{\sum_{m=-\infty}^{\infty}} g_m(r) e^{i m\phi}
\end{equation}

In terms of $\Psi(\mathbf{r})$ and $\Psi(\mathbf{\bar{r}})$ the vanishing of the divergence of $\mathbf{Q}$ reads
\begin{equation}
\label{eq:32}
\Psi(\mathbf{\bar{r}})\left[ \frac{1}{r}\frac{\partial \Psi(\mathbf{r})}{\partial r} + \frac{1}{r^2}\frac{\partial^2 \Psi(\mathbf{r})}{\partial \phi ^2} + \frac{\partial ^2\Psi(\mathbf{r})}{\partial r^2} \right] - \Psi(\mathbf{r})\left[ \frac{1}{r}\frac{\partial \Psi(\mathbf{\bar{r}})}{\partial r} + \frac{1}{r^2}\frac{\partial^2 \Psi(\mathbf{\bar{r}})}{\partial \phi ^2} + \frac{\partial^2 \Psi(\mathbf{\bar{r}})}{\partial r^2} \right] = 0
\end{equation}

\noindent
Using the representation of $\Psi(\mathbf{r})$ given in Eq.~(\ref{eq:31}) we can reformulate Eq.~(\ref{eq:32}) as follows
\begin{equation}
\label{eq:33}
\displaystyle{\sum\limits_{n,m}} \left[ \frac{1}{r} P_{n,m}(r) + \frac{1}{r^2}g_n(r) g_m(r) (n^2 - m^2) + \frac{d P_{n,m}(r)}{dr} \right] e^{i (m \phi  + n \bar{\phi})} = 0
\end{equation}
with
\begin{equation}
\label{eq:34}
P_{n,m}(r)\equiv g_n(r)\frac{d g_m(r)}{dr} - g_m(r)\frac{d g_n(r)}{dr}
\end{equation}
where we have taken into account that all the transformations we are interested in, fulfill the condition $\bar{r}=r$. In fact, $P_{n,m}(r)$ has the form of a generalized current constructed from the components of the radial part of the wave field. Furthermore, we set: $c_{n,m}(r) \equiv \frac{1}{r} P_{n,m}(r) + \frac{1}{r^2} g_n(r) g_m(r) (n^2 - m^2) + \displaystyle{\frac{d P_{n,m}(r)}{dr}}$ leading to the following relation
\begin{equation}
\label{eq:35} 
\displaystyle{\sum\limits_{n,m}} c_{n,m}(r) e^{i (m\phi  + n\bar{\phi})} = 0
\end{equation}
expressing the vanishing of the divergence of $\mathbf{Q}$.
Eq.~(\ref{eq:35}) is central for the present work since it provides the general restriction implied to the wave field $\Psi$ by the symmetry of the scattering potential. Notice that for $m=n$ it holds $P_{m,m}(r) = 0$ and therefore $c_{m,m}(r) = 0$. In addition $c_{n,m}(r)$ is antisymmetric in the indices $n,m$, i.e. $c_{n,m}(r) =  - c_{m,n}(r)$. To further utilize condition (\ref{eq:35}) we use the basis consisting of Hankel functions $H^{(\pm)}_{\ell}(r)$ to expand the radial part of the wave field $g_m(r)$ in equation (\ref{eq:31}) as follows
\begin{equation}
g_m(r)=\displaystyle{\sum_{k=-\infty}^{\infty}} \left( a_{m,k} H^{(+)}_{k}(r) + b_{m,k} H^{(-)}_{k}(r) \right)
\label{eq:36}
\end{equation}
with $a_{m,k}$, $b_{m,k}$ $\in~\mathbb{C}$.
Inserting Eq.~(\ref{eq:36}) into Eq.~(\ref{eq:35}) we obtain, after some straightforward algebraic manipulations, the constraint
\begin{multline}
\displaystyle{\sum_{n,m} \sum_{k,\ell}} \left(k^2 - \ell^2 + n^2 - m^2 \right) \left[ a_{n,\ell} a_{m,k}
H_{\ell}^{(+)}(r) H_{k}^{(+)}(r) + a_{n,\ell} b_{m,k} H_{\ell}^{(+)}(r) H_{k}^{(-)}(r) + \right. \\
\left. b_{n,\ell} a_{m,k} H_{\ell}^{(-)}(r) H_{k}^{(+)}(r) + b_{n,\ell} b_{m,k}
H_{\ell}^{(-)}(r) H_{k}^{(-)}(r) \right] e^{i (m \phi + n \bar{\phi})} = 0 \phantom{aaaaaaaaa}
\label{eq:37}
\end{multline}
holding for any $r$ and $\phi$. For the diagonal elements $n=\ell$ and $m=k$ the condition in Eq.~(\ref{eq:37}) is automatically fulfilled. We will use Eq.~(\ref{eq:37}) to determine the properties of the non-diagonal elements of the matrices $a_{m,k}$ and $b_{m,k}$ which are implied by the symmetries of the scattering potential. To illustrate how this procedure works in practice we will consider case-by-case the symmetry transforms listed in section II. The easiest way to proceed is to treat separately transformations of the form $\bar{\phi}=s  + \phi$ which are rotations from those of the form $\bar{\phi}=s- \phi$ which are reflections/inversions.

\subsection{Transformations of the form: $\bar{\phi}=s - \phi$}
In this case we find that Eq.~(\ref{eq:37}) can be written as
\begin{multline}
\displaystyle{\sum_{M,n} \sum_{k,\ell}} \left\lbrace \left(k^2 - \ell^2 - M^2 \right) \left[ \Lambda^{k,\ell}_{n,M+n}(r) - \Lambda^{k,\ell}_{n,-M+n}(r) e^{-i M s} \right] \right.\\ \left. - 2Mn \left[ \Lambda^{k,\ell}_{n,M+n}(r) + \Lambda^{k,\ell}_{n,-M+n}(r) e^{-i M s} \right] \right\rbrace e^{i n s} e^{i M \phi} = 0 \phantom{aaaaaaaaaaaaaaaaaaa}
\label{eq:38}
\end{multline}
with
\begin{multline}
\Lambda^{k,\ell}_{n,m}(r)=a_{n,\ell} a_{m,k}
H_{\ell}^{(+)}(r) H_{k}^{(+)}(r) + a_{n,\ell} b_{m,k} H_{\ell}^{(+)}(r) H_{k}^{(-)}(r)~ + \\
 b_{n,\ell} a_{m,k} H_{\ell}^{(-)}(r) H_{k}^{(+)}(r) + b_{n,\ell} b_{m,k}
H_{\ell}^{(-)}(r) H_{k}^{(-)}(r) \phantom{aaaaaaaaaaaaaaaa}
\label{eq:39}
\end{multline}
Condition (\ref{eq:38}) must hold for all $\phi$ and $r$ leading to
\begin{equation}
\Lambda^{k,\ell}_{n,M+n}(r) = \Lambda^{k,\ell}_{n,-M+n}(r) e^{-i M s}
\label{eq:40}
\end{equation}
which is fulfilled if
\begin{equation}
a_{M+n,k}=e^{-i M s} a_{-M+n,k}~~~~;~~~~b_{M+n,k}=e^{-i M s} b_{-M+n,k}
\label{eq:41}
\end{equation}
for $k \neq M+n,~-M+n$ and
\begin{equation}
\label{eq:42}
a_{N,N}=(-1)^N a_{-N,-N}~~~~~;~~~~~b_{N,N}=(-1)^N b_{-N,-N}
\end{equation}
This class of transforms consists of reflections with respect to the $x$ ($s=0$) and the $y$-axis ($s=\pi$), as well as the exchange $x \leftrightarrow y$ with $s={\pi \over 2}$. It is illuminating to present the matrix form of $a_{m,k}$, based on Eqs.~(\ref{eq:41},\ref{eq:42}) and their transposed, in each case:

\begin{itemize}

\item For $s=0$ the matrix $\mathbf{a}=a_{m,k}$ attains the form
$$\mathbf{a}=\begin{pmatrix} \dots & \dots & \dots & \dots & \dots & \dots & \dots \\
\dots & a_{-2,-2} & a_{-2,-1} & a_{-2,0} & a_{-2,-1} & a_{-2,0} & \dots \\
\dots & a_{-1,-2} & a_{-1,-1} & a_{-1,-2} & a_{-1,1} & a_{-1,-2} & \dots \\
\dots & a_{-2,0} & a_{-2,-1} & a_{0,0} & a_{-2,-1} & a_{-2,0} & \dots \\ 
\dots & a_{-1,-2} & a_{-1,1} & a_{-1,-2} & -a_{-1,-1} & a_{-1,-2} & \dots \\
\dots & a_{-2,0} & a_{-2,-1} & a_{-2,0} & a_{-2,-1} & a_{-2,-2} & \dots \\
\dots & \dots & \dots & \dots & \dots & \dots & \dots 
\end{pmatrix} $$

\item The $\mathbf{a}$-matrix for $s=\pi$ becomes
$$\mathbf{a}=\begin{pmatrix} \dots & \dots & \dots & \dots & \dots & \dots & \dots \\
\dots & a_{-2,-2} & a_{-2,-1} & -a_{-2,0} & -a_{-2,-1} & a_{-2,0} & \dots \\
\dots & a_{-1,-2} & a_{-1,-1} & -a_{-1,-2} & -a_{-1,1} & a_{-1,-2} & \dots \\
\dots & -a_{-2,0} & -a_{-2,-1} & a_{0,0} & a_{-2,-1} & -a_{-2,0} & \dots \\ 
\dots & -a_{-1,-2} & -a_{-1,1} & a_{-1,-2} & -a_{-1,-1} & -a_{-1,-2} & \dots \\
\dots & a_{-2,0} & a_{-2,-1} & -a_{-2,0} & -a_{-2,-1} & a_{-2,-2} & \dots \\
\dots & \dots & \dots & \dots & \dots & \dots & \dots 
\end{pmatrix} $$

\item Finally, for $s={\pi \over 2}$ the matrix $a_{m,k}$ is written as
$$\mathbf{a}=\begin{pmatrix} \dots & \dots & \dots & \dots & \dots & \dots & \dots \\
\dots & a_{-2,-2} & a_{-2,-1} & a_{-2,0} & i a_{-2,-1} & i a_{-2,0} & \dots \\
\dots & a_{-1,-2} & a_{-1,-1} & i a_{-1,-2} & a_{-1,1} & -a_{-1,-2} & \dots \\
\dots & a_{-2,0} & i a_{-2,-1} & a_{0,0} & -a_{-2,-1} & -a_{-2,0} & \dots \\ 
\dots & i a_{-1,-2} & a_{-1,1} & -a_{-1,-2} & -a_{-1,-1} & -i a_{-1,-2} & \dots \\
\dots & ia_{-2,0} & -a_{-2,-1} & -a_{-2,0} & -i a_{-2,-1} & a_{-2,-2} & \dots \\
\dots & \dots & \dots & \dots & \dots & \dots & \dots 
\end{pmatrix} $$

\end{itemize} 
Similar expressions hold also for $b_{m,k}$. Notice that a non-diagonal element of the matrix $a_{m,k}$ (or $b_{m,k}$) is obtained from the non-diagonal element of the same column and the pre-previous row (or the same row and the pre-previous column) through multiplication by $i$. In fact the off-diagonal entries of the matrices $a_{m,k}$, $b_{m,k}$ are in one-to-one relation to the off-diagonal elements of the corresponding $\mathbf{S}$-matrix. 

\subsection{Transformations of the form: $\bar{\phi}=s + \phi$}

For this class of transformations Eq.~(\ref{eq:37}) leads to the general result that the off-diagonal matrix elements of $a_{m,k}$ and $b_{m,k}$ are non-zero only for $m$-values differing by $\Delta m= \frac{2 \pi \nu}{s}$ with $\nu$ $\in~\mathbb{Z}$. Thus, the matrices $\mathbf{a}$ and $\mathbf{b}$ are in general sparse becoming sparser as the angle $s$ becomes smaller. Assuming that $s={\pi \over q}$ then we observe that the matrices $\mathbf{a}$, $\mathbf{b}$ are sparser whenever $q$ is closer to an irrational number becoming purely diagonal for $q$ irrational. A simple example is provided by the case of scattering off a potential which is symmetric under rotation by $s={\pi \over 2}$. In such a case only rows differing by $4$ in their row-index have non-vanishing entries and therefore the matrix $\mathbf{a}$ attains the form

$$\mathbf{a}=\begin{pmatrix} \dots & \dots & \dots & \dots & \dots & \dots & \dots \\
\dots & a_{-n,-n} & a_{-n,-n+1} & a_{-n,-n+2} & a_{-n,-n+3} & a_{-n,-n+4} & \dots \\
\dots & 0 & a_{-n+1,-n+1} & 0 & 0 & 0 & \dots \\
\dots & 0 & 0 & a_{-n+2,-n+2} & 0 & 0 & \dots \\
\dots & 0 & 0 & 0 & a_{-n+3,-n+3} & 0 & \dots \\
\dots & a_{-n+4,-n} & a_{-n+4,-n+1} & a_{-n+4,-n+2} & a_{-n+4,-n+3} & a_{-n+4,-n+4} & \dots \\
\dots & \dots & \dots & \dots & \dots & \dots & \dots 
\end{pmatrix} $$

\noindent
and similarly for the matrix $\mathbf{b}$. In the limit of continuous rotational symmetry only the diagonal elements are non-zero and the usual ansatz presented in Eqs.~(\ref{eq:29}) and (\ref{eq:30}) is recovered. These considerations complete the cases of linear symmetry transformations  leaving the scattering potential invariant as described in section II. The derived conditions for the non-diagonal entries of the radial matrices $\mathbf{a}$ and $\mathbf{b}$ provide strong constraints, dictated by the symmetries of the scattering potential, on the transition probabilities between different radial modes for given angular momentum. They reduce significantly the number of coefficients in the general expansion of the wave field with respect to the 2-d basis in polar coordinates and can simplify solving the scattering problem under consideration based on numerical methods. Clearly, our approach is general and can be also applied to other classes of symmetry transforms according to the description in \cite{Diakonos2019}. However, such a study goes beyond the scope of the present work.

\section{Summary and conclusions}\label{sec:5}

In the present work we have used the divergence free non-local currents derived in \cite{Kalozoumis2014,Spourdalakis2016,Diakonos2019} to obtain symmetry induced general properties  of the wave field for two-dimensional wave scattering. Since the scattering boundary conditions usually break the symmetry present in the Hamiltonian of the considered system, the impact of the symmetries of the scattering potential on the form of the wave field is not obvious at all. The common strategy is to consider the problem in the asymptotic far-field regime, where informations concerning the symmetry of the scattering potential -in particular for finite support potentials- are lost. In contrary, the approach presented here is exact, taking into account also near field properties. Thus, our analysis reveals the role of the symmetry induced non-local currents as an invaluable tool to achieve a consistent quantitative description of scattering in higher dimensional problems based on first principles. The conditions derived here are generally applicable providing a useful guide for solving 2-d scattering problems in the presence of symmetries. They determine the relations between transition probabilities connecting different radial modes for a given angular momentum. Here we have restricted our study to two-dimensional problems involving specific scattering potential symmetries. Nevertheless, the extension to higher dimensions and a wider class of symmetry transforms is possible and will be addressed in a future work.    

{}


\begin{thebibliography}{99}

\bibitem{book1} J.~R. Taylor, {\it "Scattering Theory: The Quantum Theory of Nonrelativistic Collisions"},
R.E. Krieger Publishing Company, 1972.

\bibitem{book2} R.~G. Newton, {\it "Scattering Theory of Waves and Particles"}, Dover Books on Physics, 1982.

\bibitem{scatt_sym} Y. Alhassid, F. Gursey and F. Iachello, Ann. Phys. {\bf 148},
346 (1983); Y. Alhassid, F. Gursey and F. Iachello, Phys. Rev. Lett. {\bf 50}, 873 (1983); A. Frank and K.~B. Wolf, Phys. Rev. Lett. {\bf 52}, 1737 (1984); C.~W. Lee and U. Fano, Phys. Rev. A {\bf 36}, 66 (1987); Y. Wu, J. Math. Phys. {\bf 28}, 1360 (1987); A.~I. Nivishov, Theor. Math. Phys. {\bf 98}, 42 (1994); P. Pereyra, J. Math. Phys. {\bf 36}, 1166 (1995); U. Fano, D. Green, J.~L. Bohn and T.~A. Heim, J. Phys. B {\bf 32}, R31 (1999); E. de Prunel\'{e}, J. Math. Phys. {\bf 59}, 102102 (2018); A.~G. Ramm, Appl. Math. Lett. {\bf 96}, 122 (2019).

\bibitem{Asymp} I.~R. Lapidus, Am. J. Phys. {\bf 50}, 45 (1982); S.~K. Adhikari, Am. J. Phys. {\bf 54}, 362 (1986); M. Rosenkranz and W. Bao, Phys. Rev. A {\bf 84}, 050701(R) (2011); E.~A. Koval, O.~A. Koval and V.~S. Melezhik, Phys. Rev. A {\bf 89}, 052710 (2014); S. McAlinden and J. Shertzer, Am. J. Phys. {\bf 84}, 764 (2016); L. Zhang, J. Acoust. Soc. Am. {\bf 145}, EL185 (2019).

\bibitem{Ramm} A.~G. Ramm, J. Math. Anal. Appl. {\bf 156}, 333 (1991). 

\bibitem{Liu2014} T. Liu, W.-D. Li, and W.-S. Dai, JHEP {\bf 06}, 087 (2014);  W.-D. Li, and W.-S. Dai, J. Phys. A {\bf 49}, 465202 (2016).

\bibitem{Kalozoumis2014} P.~A. Kalozoumis, C. Morfonios, F.~K. Diakonos, and P. Schmelcher, Phys. Rev. Lett. {\bf 113}, 050403 (2014).

\bibitem{Kalozoumis2015} P.~A. Kalozoumis, C.~V. Morfonios, F.~K. Diakonos, and P. Schmelcher, Ann. Phys.(NY) {\bf 362}, 684 (2015).

\bibitem{LocSym} P.~A. Kalozoumis, C. Morfonios, F.~K. Diakonos, and P. Schmelcher, Phys. Rev. A {\bf 87}, 032113 (2013); P.~A. Kalozoumis, C. Morfonios, N. Palaiodimopoulos, F.~K. Diakonos, and P. Schmelcher, Phys. Rev. A {\bf 88}, 033857 (2013); P.~A. Kalozoumis, G. Pappas, F.~K. Diakonos, and P. Schmelcher, Phys. Rev. A \textbf{90}, 043809 (2014); C. Morfonios, P. Schmelcher, P.~A. Kalozoumis, and F.~K. Diakonos, Nonlin. Dyn. {\bf 78}, 71 (2014); P.~A. Kalozoumis, O. Richoux, F.~K. Diakonos, G. Theocharis, and P. Schmelcher, Phys. Rev. B {\bf 92}, 014303 (2015);  V.~E. Zampetakis, M.~K. Diakonou, C.~V. Morfonios, P.~A. Kalozoumis, F.~K. Diakonos, and P. Schmelcher, J. Phys. A: Math. and Theor. {\bf 49}, 195304 (2016).

\bibitem{Diakonos2019} F.~K. Diakonos and P. Schmelcher, J. Phys. A: Math. and Theor. {\bf 52}, 1552034 (2019).

\bibitem{Schmelcher2017} P. Schmelcher, S. Kr\"{o}nke and F.~K. Diakonos, J. Chem. Phys. {\bf 146}, 044116 (2017).

\bibitem{Spourdalakis2016} A.~G.~B. Spourdalakis, G. Pappas, C.~V. Morfonios, P.~A. Kalozoumis, F.~K. Diakonos and P. Schmelcher, Phys. Rev. A {\bf 94}, 052122 (2016).

\bibitem{Metaxas} M. Metaxas, P. Schmelcher and F.K. Diakonos, in preparation.

\bibitem{Barbarosie2011} C. Barbarosie, Quaterly of App. Math. {\bf LXIX}, 309 (2011).

\bibitem{Abramowitz1973}  M. Abramowitz and I.~A. Stegun, 
\textit{"Handbook of Mathematical Functions: with Formulas, Graphs, and Mathematical Tables"}, Dover Pub., 1973.



\end{thebibliography}
\end{document}